\documentclass[11pt,a4paper]{article}
\usepackage{amsmath}
\usepackage{amssymb}
\usepackage{amsfonts}
\usepackage{amstext}
\usepackage{amsbsy}
\usepackage[mathscr]{eucal}
\usepackage{graphicx}
\usepackage{color}
\usepackage[all]{xy}

\renewcommand{\phi}{\varphi}

\newcommand{\ket}[1]{|{#1}\rangle}

\newcommand{\obs}[1]{\hat{#1}}
\newtheorem{thm}{Theorem}[subsection]
 \newtheorem{cor}[thm]{Corollary}

 \newtheorem{prop}[thm]{Proposition}
 \newtheorem{rmk}[thm]{Remark}
 \numberwithin{equation}{subsection}

\makeindex
\title{Geometric characterization of mixed quantum states}

\author{\textit{ Hoshang Heydari}\\
\small\textit{Department Of Physics, University of Isfahan, Isfahan, Iran}
}

\date{}
\begin{document}
\maketitle \thispagestyle{empty} \maketitle 

\begin{abstract}
Characterization of mixed quantum states represented by density operator is one of the most important task in quantum information processing. In this work we will present a geometric approach to characterize the density operator in terms of fiber bundle over a quantum phase space. The geometrical structure of the quantum phase space
of an isospectral mixed quantum states can be realized as a co-adjoint orbit of a Lie group equipped with a specific K\"{a}hler structure.
 In particular we will  briefly discuss the construction of a fiber bundle over the quantum phase space based on symplectic reduction and purification method. We will also show that the map is a Riemannian submersion which enable us to provide some applications of the geometric framework such as geometric phase and quantum speed limit.
\end{abstract}

\section{Introduction}
In geometric quantum mechanics, the systems are described based on their underlying geometrical structures \cite{Gunter_1977,Kibble_1979,Ashtekar_etal1998,Brody_etal1999}. Recently, it has been shown that such geometrical structures of quantum theory provide us with useful information on the foundations of the theory with many applications in quantum science and technology \cite{Zanardi_etal1999,Solinas_etal2003,Uhlmann1989,Uhlmann1991}.
In geometric formulation of  quantum mechanics, the projective Hilbert space is constructed by general Hopf fibration of a hypersphere and usually is called the quantum phase space of a pure quantum state. However, pure quantum states are a small subclass of all quantum states. On the other hand, mixed quantum states represented by density operators $\rho$ are the most general  states in quantum mechanics.
\\

In this work we will show that the quantum phase space of  mixed quantum states is a  generalized flag manifold which is equipped with a symplectic form that also coincides with a specific K\"{a}hler structure which is called Kirillov-Kostant-Souriau K\"ahler (KKS) symplectic form on the co-adjoint orbit  and a Riemannian metric \cite{Hosh}. Then we will construct a fiber bundle over the quantum phase space and show that the map is a Riemannian submersion \cite{Mont,GUR}. Finally we will briefly discuss some applications of fiberbundle  such as a geometric phase based on the holonomy group and a quantum speed limit for unitary evolving mixed quantum states. Please note that recently we have introduced a geometric framework for unitary evolving mixed quantum states based on purification bundle and momentum mapping \cite{GUR}. In that framework the metric on the total space (Hilbert space) is the restriction of real part of  Hilbert-Schmitd inner-product. In current work the metric on the total space is the Killing form of a unitary group. These two metric are different. The advantages with the current geometric framework based on the co-adjoint orbit and the generalized flag manifold is that we have an explicit expression for the almost complex structure of quantum phase space and the Lie groups that  we consider here do have rich mathematical structures which enable us to investigate their applications in quantum information and quantum computation.

\section{Geometric framework}\label{s1}

 In this section we will review our geometric framework for mixed quantum states based on  KKS structure.
First,  we will review the basic definition and construction of the mixed quantum states based on a k\"{a}hler structure \cite{Hosh}. Then,  we will discuss symplectic form and an almost complex structures on the quantum phase space.
Let
\begin{equation}
\mathrm{Her}(\mathcal{H})=\{A\in M_{n}(\mathbb{C}):A^{\dagger}=A\}
\end{equation}
be the space of Hermitian matrices. Then the orbit of $U(n)$ action on $\mathrm{Her}(\mathcal{H})$ is the submanifolds with real eigenvalues which are diagonalizable
\begin{equation}
\mathcal{D}(\sigma)=\{\rho\in \mathrm{Her}(\mathcal{H}): ~\text{with}~\text{spectrum}~ \sigma=(p_{1},p_{2},\ldots,p_{k})\in \mathbb{R}^{n}\},
\end{equation}
that is $\rho=UDU^{-1}$, where $D$ is diagonal and $U\in U(n)$. Thus the coadjoint of $U(n)$ are the same as conjugacy classes of $D$.
\begin{prop}
Let $\sigma=(p_{1},p_{2},\ldots,p_{k})$ consists of $k$ distinct values with multiplicities $n_{1},n_{2},\ldots,n_{k}$. Then the orbit space $\mathcal{D}(\sigma)$ is diffeomorphic with homogeneous space $U(n)/U(n_{1})\times U(n_{2})\times\cdots \times U(n_{k})\cong SU(n)/S(U(n_{1})\times U(n_{2})\times\cdots \times U(n_{k}))$.
\end{prop}
\begin{cor}
If $p_{1}>p_{2}=\cdots=p_{k}$, then $\mathcal{D}(\sigma)$ is diffeomorphic to  $\mathbb{C}P^{n-1}$.
\end{cor}
proof follows from the observation that   a hermitian matrix with spectrum $\sigma$ is completely determined by its $p_{1}-$eigenspace.
\\
 The quantum phase space can be equipped by a hermitian inner-product
\begin{equation}
h(X,Y)=g(X,Y)+i \tilde{\omega}(X,Y),
\end{equation}
where $g(X,Y)=\tilde{\omega}(X,JY)$ is a Riemannian metric on quantum phase space \cite{Hosh}.
The importance of this form stems from the fact that
if $A$ is the expectation value function of a Hermitian operator $\obs{A}$, that is
\begin{equation}
A(\rho)=\mathrm{Tr}(\rho \obs{A}),
\end{equation}
and
$X_A$ is the Hamiltonian vector field associated with $A$, which is implicitly defined by the identity $dA(X)=\omega(X_A,X),$ then
\begin{equation}
X_A(\rho)=\frac{1}{i\hbar}[\hat{A},\rho].
\end{equation}
Thus, in particular, the von Neumann equation on $\mathcal{D}(\sigma)$ is the Hamiltonian flow equation of the expected energy function with respect to the symplectic form.

\section{Fiber bundle structure of the quantum phase space}\label{subsec5}

In this section we will attempt to construct a fiber bundle over the quantum phase space. In particular, first  we will show that the map from $U(n)$ to $ U(n_{1}\times U( n_{2})\times \cdots\times U( n_{k})=U(\sigma)$ is a Riemannian submersion and then construct a fiber bundle over the quantum phase space.

\subsection{Riemannian submersion }

 A reductive homogeneous spaces has a fixed decomposition of the Lie algebra $\mathfrak{u}(n)=\mathfrak{h}\oplus \mathfrak{m}$ such that $\mathrm{Ad}(\mathfrak{h})\mathfrak{m}\subset \mathfrak{m}$, where $\mathfrak{h}=\mathfrak{u}(n_{1})\oplus \mathfrak{u}(n_{2})\oplus\cdots\oplus \mathfrak{u}(n_{k})$.
Moreover, a homogeneous space is called  reductive if there exists a decomposition of Lie algebra $\mathfrak{u}(n)=\mathfrak{h}\oplus \mathfrak{m}$ such that
\begin{equation}
A\mathrm{d}(U(\sigma)) \mathfrak{m}\subset \mathfrak{m}.
\end{equation}
\begin{figure}[htbp]\label{fig11}
\centering
\includegraphics[width=0.7\textwidth,height=0.3\textheight]{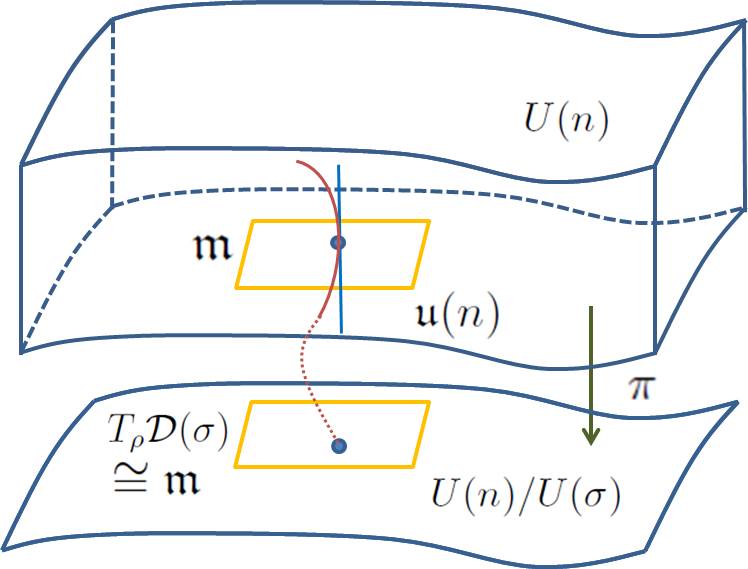}
\caption{Illustration of the bundle $\pi$ and the decomposition of $T_{\psi}U(n)$.}
\label{bundle}
\end{figure}
Let the  map $\pi: U(n)\longrightarrow \mathcal{D}(\sigma)=U(n)/U(\sigma)$ be a natural projection with $\rho=\psi U(\sigma)\in \mathcal{D}(\sigma)$. Then
\begin{equation}
\pi_{*\psi}: T_{\psi}U(n)\longrightarrow T_{\rho} \mathcal{D}(\sigma)
\end{equation}
will
induce an isomorphism
\begin{equation}
 T_{\rho} \mathcal{D}(\sigma)\cong T_{\psi}U(n)\cong \mathfrak{u}(n)/\mathfrak{h}\cong \mathfrak{m}.
\end{equation}
The map  $\pi: U(n)\longrightarrow \mathcal{D}(\sigma)=U(n)/U(\sigma)$ is a fibration and also a submersion.
\begin{rmk}
To make sure that the Killing form is non-degenerate on $\mathfrak{m}$, we could choose that the total space be the Lie group $SU(n)$, but then we realized that the group $U(n)$ also satisfies this condition since $U(n)$ is compact Lie group. This will make the construction of the fiberbundle much easier which we will consider in next section.
\end{rmk}
Now, he Killing form $B(X,Y)$ of $U(n)$ is non-degenerate on $\mathfrak{m}$, thus the symplectic form  $\omega$ can be written uniquely in terms of $B$-antisymmetric linear transformation $\varphi:\mathfrak{m}\longrightarrow\mathfrak{m}$ on $\mathfrak{m}$ such that
\begin{equation}
\omega(X,Y)=B(\varphi(X),Y).
\end{equation}
One can show that for a closed form $\omega$, $\varphi$ can be written as $ad(\gamma)$ for all $\gamma\in \mathfrak{u}(n)$. In this case we get
\begin{equation}
\omega(X,Y)=B([\gamma,X],Y)=B(\gamma,[X,Y]).
\end{equation}
Let the isotropy $U(\sigma)$ of the  manifold $U(n)/U(\sigma)$  be represented as block diagonal so that
in  the case of $U(\sigma) $ maximal torus (hence largest dimensional co-adjoint orbit),
$U(\sigma)$ consists of the diagonal matrices.
 The tangent bundle $TU(n)$ can be decomposed to vertical $VU(n)$ sub bundle and a horizontal sub bundle $HU(n)$, that is
\begin{equation}
TU(n)=HU(n)\oplus VU(n),
\end{equation}
where $VU(n)=\mathrm{Ker} d\pi$ and $HU(n)=VU(n)^{\perp}$ with $\perp$ be the orthogonal complement with respect to $U(n)$. Now,  a Riemannian submersion $\pi: U(n)\longrightarrow \mathcal{D}(\sigma)=U(n)/U(\sigma)$ is a submersion with the property that $d_{\psi}\pi$ is an isometry  when restricted to $HU(n)=(\mathrm{Ker} d\pi)^{\perp}$.\\
Let $E_\alpha$ be a Killing  orthonormal basis for their Killing-orthogonal complement.  So, in the case $U(\sigma)$ is a diagonal matrix, this would mean that the  basis $E_\alpha$  for the Lie algebra $\mathfrak{u}(n)$  are generated by $\frac{1}{\sqrt{2}}(e^{ij}-e^{ji})$ for $1\leq i\leq j\leq n$, where $e^{ij}$ is the matrix with 1 in position $(i,j)$ and zero elsewhere.
(We could multiply it by $\sqrt{-1}$ to make the model a skew-Hermitian rather than Hermitian).
These same $E_\alpha$ are a basis for the tangent space $T\mathcal{D}(\sigma)$ to $U(n)/U(\sigma)$.
The induced-by-Riemannian submersion metric for $U(n)/U(\sigma)$ is the same.
Namely the $E_\alpha$ continue to be orthonormal.
This means that we have
\begin{equation}g(E_\alpha, E_\beta) = \delta_{\alpha\beta}.
\end{equation}
Thus we have a Riemannian submersion $U(n)\longrightarrow U(n)/U(\sigma)$, where the  bi-invariant metric on $U(n)$ is generated by the Killing form $B(X,Y)$ defined by
   \begin{equation}
   B(X,Y)=\mathrm{Tr}(adX\circ adY)
  \end{equation}
   and the symplectic form is the same as the KKS form $\omega_{KKS}$.
    \begin{rmk}
    For the Lie algebra $\mathfrak{u}(n)$, we have:(i) if $X\in u(n)$, then $X^{\dagger}\in u(n)$, (ii) $\mathrm{Tr}(XY)$ is real for all $X,Y\in u(n)$ which indicate that the trace is non-degenerate, (iii)
   $B(X,Y)= c\mathrm{Tr}(XY)$ with $c\neq0$, that is the Killing form is proportional to the trace form \cite{ONeill}. We can e.g., choose $c=1/2$, which corresponds to a natural Hermitian product on $\mathbb{C}^{n}$.
  \end{rmk}
   Moreover on $U(n)/U(\sigma)$ the metric is $g_{KKS}(X,Y)=\omega_{KKS}(X,JY)$. Note also that $B(X,Y)$ is $Ad$-invariant, that is $B(X,Y)=B(Ad(U)X,Ad(U)Y)$ for all $U\in U(n)$ and $X,Y\in \mathfrak{u}(n)$.
\\
Consider a density operator with $k$- dimensional support having a spectrum given by $\sigma=(p_{1},p_{2},\ldots,p_{k})$, where $p_{i}$, for all $i=1,\ldots, k$ are positive eigenvalues listed in descending order. Moreover, we let $\mathbb{C}^{k}$ be a $k$-dimensional Hilbert space which is spanned by the following  orthonormal basis $\{\ket{i}\}^{k}_{i=1}$. Furthermore, let $\mathcal{L}(\mathbb{C}^{k},\mathcal{H})$ be the space of linear operator from $\mathbb{C}^{k}$ to $\mathcal{H}$. Next we could identify the $\mathcal{D}(\sigma)$ as $U(n)$-orbit with a fix spectrum $\sigma$.
Now, the map  \begin{equation}
\pi:U(n)\longrightarrow U(n)/U(\sigma),
\end{equation}
is defined by
$\pi(\psi)=\psi\psi^{\dagger}$ is a principal fiber bundle with right acting group
 $U(\sigma)=U(n_{1})\times U( n_{2})\times \cdots\times U( n_{k})$, where $\psi\in \mathcal{L}(\mathbb{C}^{k},\mathcal{H})$. We leave the detail construction of the purification bundle over co-adjoint orbit here and refer the interested reader to our recent work \cite{GUR}. But we will emphasis that these two geometric frame work for mixed quantum states are different and could have different applications.

\section{Some applications }
In this section we briefly discuss some applications of our geometric framework such as geometric phase and quantum speed limit for unitary evolving mixed quantum states. In order to define a geometric phase we need to define a connection on tangent bundle of $U(n)$ as follows. Let
\begin{equation}
\mathcal{A}:TU(n)\longrightarrow \mathfrak{u}(\sigma)
\end{equation}
defined by  $\mathcal{A}_{\psi}=\mathbb{I}^{-1}_{\psi}\mathbb{J}_{\psi}$, where
$\mathbb{J}:TU(n)\longrightarrow \mathfrak{u}(\sigma)^{*}$ is metric momentum map and $\mathbb{I}:TU(n)\times \mathfrak{u}(\sigma)\longrightarrow \mathfrak{u}(\sigma)^{*}$ is the locked inertia tensor defined by
 $\mathbb{J}_{\psi}(X)(\xi) =B(X, \hat{\xi}(\psi))$ and $\mathbb{I}_{\psi}\xi(\eta)=B(\hat{\xi}(\psi),\hat{\eta}(\psi))$ respectively. $\mathcal{A}_{\psi}$ is called a mechanical connection. Next we define a parallel transport operator $\Pi(\rho)$ from the fibre over $\rho(0)=\rho_{0}$ onto the fibre $\rho(\tau)$ in terms of the mechanical connection \cite{GP}. Now the geometric phase of $\rho$ is defined by
\begin{eqnarray}
\gamma_{g}(\rho)&=&\arg \mathrm{Tr}(P(\sigma)\mathrm{Hol}(\rho))
\\\nonumber&=&
\arg \mathrm{Tr}(\psi^{\dagger}_{0}\Pi[\rho]\psi_{0})
\end{eqnarray}
where $\mathrm{Hol}(\rho)$ is the holonomy of $\rho$.
Next we will discuss another application of our geometric framework which is very important in quantum information processing namely the quantum speed limit.
First we note that the geodesic distance between two isospectral density operators is the minimum length of all curve in the quantum phase space. Since we have shown that our bundle $\pi$ is a Riemannian submersion then all horizontal lifting curves is length preserving. This implies that a curve in the quantum phase space is a geodesics if and only if its horizontal lift is a geodesics in $U(n)$.  A real-valued  function $H:\mathcal{H}\longrightarrow \mathbb{R}$ of  $\hat{H}$ is called average energy function and it is defined by $H(\rho)=\mathrm{Tr}(\hat{H}\rho)$. If we let $X_{H}$ denotes the Hamiltonian vector field of $H$, then the von Neumann equation can be written as
\begin{equation}
X_{H}(\rho)=\frac{1}{i\hbar}[\hat{H},\rho].
\end{equation}
The Hamiltonian vector field has a gauge-invariant lift $X_{\hat{H}}$ to $U(n)$ which is defined by
  \begin{equation}
X_{\hat{H}}(\psi)=\frac{1}{i\hbar}\hat{H}\psi.
\end{equation}
Next, for a given Hamiltonian, we will establish a relation between the uncertainty function
   \begin{equation}
\Delta H(\rho)=\sqrt{\mathrm{Tr}(\hat{H}^{2}\rho)-\mathrm{Tr}(\hat{H}\rho)^{2}},
\end{equation}
and the metric, that is the Hamiltonian vector field $X_{H}$ satisfies
   \begin{equation}
\hbar^{2} g_{KKS}(X_{H}(\rho),X_{H}(\rho))\geq\Delta H(\rho)^{2}.
\end{equation}
If the  Hamiltonian $\hat{H}$ is parallel at $\rho$, then $\hbar^{2} g_{KKS}(X_{H}(\rho),X_{H}(\rho))=\Delta \hat{H}(\rho)^{2}$. Note that the Hamiltonian $\hat{H}$ is  parallel at a density operator $\rho$ if $X_{\hat{H}}(\psi)$ horizontal at every $\psi$ in the fiber over $\rho$.
 Now, let
 $\rho_{0},\rho_{1}\in \mathcal{D}(\sigma)$ be two density operators and $\hat{H}$ be the Hamiltonian of a quantum system. Then distance between $\rho_{0}$ and $\rho_{1}$ is given by
\begin{equation}
\mathcal{D}(\rho_{0},\rho_{1})\leq\frac{1}{\hbar}\int^{t=\tau}_{t=0}
\Delta H(\rho) dt,
\end{equation}
where  $\rho(0)=\rho_{0}$ and $\rho(\tau)=\rho_{1}$.
Now, we are able to give a quantum speed limit for unitary evolving mixed quantum state based on K\"{a}hler structure as follows. Let
\begin{equation}
\Delta E=\frac{1}{\tau}\int^{\tau}_{0}\Delta H(\rho)\mathrm{d}t.
\end{equation}
Then, we have the following geometric quantum speed limit
\begin{equation}
\tau\leq \frac{\hbar}{\Delta E} \mathcal{D}(\rho_{0},\rho_{1}).
\end{equation}
 Note that our new geometric quantum  distance measure  always measure a shorter distance than quantum distance measure $Dist(\rho_1,\rho_1)$ that  we have constructed in our recent paper \cite{QSL}.
Thus we have
\begin{equation}
\tau\geq \frac{\hbar}{\Delta E} Dist(\rho_1,\rho_1)\geq \frac{\hbar}{\Delta E} \mathcal{D}(\rho_{0},\rho_{1}).
\end{equation}

\section{Conclusion}
In this paper, we have investigated the geometrical structure of mixed quantum states. After reviewing the construction of quantum phase space based on the co-adjoint method, we have identified the quantum phase space with a reductive  homogeneous space. The main result of the paper is the construction of a  fiber bundle over the quantum phase space based on the symplectic reduction. We have also briefly discussed some applications of the geometric framework such as geometric phase and geometric quantum speed limit. In particular we  have considered the construction of a geometric phase based on the holonomy group which is defined by the connection on the total space of the fiber bundle and derived a geometric speed limit based on a dynamical distance on the quantum phase space. In the continuation of this geometric framework for mixed quantum states, we will investigate many possible applications in the field of quantum information and quantum computation.

\begin{flushleft}
\textbf{Acknowledgments:} The author acknowledges useful  email conversations with Professor R. Montgomery.   The author also acknowledges useful comments and discussions with Professor R. Roknizadeh and Ole Andersson.
\end{flushleft}

\end{document}